\documentclass{gGAF2e}

\usepackage{amsbsy,amssymb,amsmath,bm}
\usepackage{graphicx,subfigure}
\usepackage[authoryear]{natbib}

\begin{document}

\markboth{\rm X. WEI}{\rm GEOPHYSICAL \&  ASTROPHYSICAL FLUID DYNAMICS}

\title{On the evolution of global ocean tides}
\author{Xing Wei\thanks{Email: xingwei@bnu.edu.cn} \\ Department of Astronomy, Beijing Normal University, Beijing 100875, China}

\maketitle

\begin{abstract}
We apply Laplace's tidal theory to the evolution of lunar and solar tides on the geologic timescale of Earth's rotation and focus on the tidal resonance. We study the global tide in the mid-ocean far away from continents. On the short timescale, a linear relationship of tidal height and Earth's rotation is obtained. On the long timescale, the tide is less than 1 metre at present but it was 5 metres in the past and will reach 8 metres in the future because of resonances of tidal wave and Earth's rotation. We conclude that the Earth-Moon orbital separation and the slowdown of Earth's rotation are faster than expected before.
\end{abstract}


\section{Introduction}\label{intro}

The theory of ocean tide can be dated back to Newton who formulated the tidal force, i.e. the gravitational perturbation exerted by Moon or Sun on Earth, which is inversely proportional to the cube of distance. Moon's tidal contribution is larger than Sun's because Moon is much closer to Earth though its mass is much smaller. Newton's theory is static and in 1775 Laplace established a dynamic theory using a thin ocean layer covering on Earth. In his theory Earth's rotation and the fluid motion near Earth's surface were considered, and tidal wave was found \citep{souchay}. Laplace's theory is mathematically described with Laplace's tidal equations, i.e. a two-dimensional shallow water model \citep{pedlosky}. In Laplace's dynamic theory, gravity and rotation are the two major factors to influence tidal wave. Consequently, tidal wave is a mixed mode of surface gravity wave, e.g. ripples on lake surface \citep{lighthill}, and inertial wave, e.g. fluid column on top of an obstacle in a rotating tank \citep{greenspan}. In astronomy, the former is called $f$ mode and the latter $r$ mode \citep{goodman2009}. Tidal wave is dynamical tide and differs from equilibrium tide which is a result of hydrostatic balance \citep{ogilvie2014}. Tidal resonance can occur when tidal frequency is close to the natural frequency of tidal wave \citep{zahn1970, goldreich1989, wei2019}. The observation of tidal rhythmites shows that the present Moon's recession rate is higher than that millions of years ago \citep{williams}, which implies that Earth is approaching tidal resonance.

Local geography, e.g. continental shorelines and oceanic basins, can greatly influence ocean tide. This effect has been extensively studied since 1970s \citep[e.g.,][]{platzman}. On the other hand, the global tide in the mid-ocean far away from continents evolves on the geologic timescale of Earth's rotation and tidal resonances can occur on this timescale. Take the Earth-Moon system for example. The total angular momentum conserves but the total energy does not. The energy dissipation arises mainly from the friction near the rough seafloors and pushes the Earth-Moon system to the equilibrium state of minimum energy, and eventually both Earth's and Moon's rotations will synchronise with the orbit. Because the total angular momentum deposits mainly in the orbit, the orbital period varies not too much. If the total angular momentum conservation is precisely calculated, i.e. the orbital angular momentum together with the Earth's and Moon's rotational angular momenta, then it is indeed that the orbital period slows down and the separation between Earth and Moon increases \citep{bills1999}. However, in the present study, we do not consider the mutual interaction of orbital dynamics and tidal friction but focus on the tidal resonance, i.e. how strong the global ocean tide can reach when tidal resonances occur. Moon's rotation has already reached the synchronisation state because of its small mass, and Earth's rotation slows down on a geologic timescale towards the synchronisation state. The present period of Earth's rotation is one day, but it was several hours when Earth formed in the past, e.g. 5 hours \citep{tethered-moon} or 6 hours after giant impact \citep{impact}, and will reach about one month in the future. The observation shows that the mid-ocean experiences tide of 1 metre or less \citep{thurman}. Tide can be higher in the past and future when Earth's rotation is appropriate to induce tidal resonances.

For the present Earth, the rotational frequency $\varOmega$ is much lower than the dynamical frequency $\sqrt{g/R}$ where $R$ is radius and $g$ is surface gravity. In the regime $\varOmega/\sqrt{g/R} \approx 0.06 \ll 1$ the free-oscillation problem in the absence of tidal force has been asymptotically studied by \citet{longuet-higgins}. The forced-oscillation problem was also qualitatively discussed in the same paper but the exact tidal height was not given because rotation couples different spherical harmonics and it is difficult to do analytical calculation in this case. In \citet{webb1980} the tidal resonance was numerically investigated and the powers of different spherical harmonics of tidal waves were obtained. However, the resonance between tidal wave and Earth's rotation and the evolution of global tide on the geologic timescale of Earth's rotation has not been well studied. In the present paper, we will numerically study the global ocean tide in the past and future when Earth's rotation varies in a very wide regime of $\varOmega$. In Section \ref{model} the model is established. In Section \ref{results} the results are discussed. In Section \ref{summary} the concluding remarks are given.

\section{Model}\label{model}

Suppose that a thin fluid layer with uniform depth covering on the surface of a rotating body. We apply Laplace's tidal equations in the frame rotating with Earth \citep{souchay} by adding a frictional force of Rayleigh drag proportional to velocity
\begin{subequations}
\label{dimensional}
\begin{align}
\frac{\upartial u_\theta}{\upartial t}-2\varOmega\cos\theta u_\phi=&\,\frac{1}{R}\frac{\upartial}{\upartial\theta}(-g\eta+V+\varPsi)-\gamma u_\theta, \\
\frac{\upartial u_\phi}{\upartial t}+2\varOmega\cos\theta u_\theta=&\,\frac{1}{R\sin\theta}\frac{\upartial}{\upartial\phi}(-g\eta+V+\varPsi)-\gamma u_\phi, \\
 \frac{\upartial\eta}{\upartial t}+\frac{h}{R\sin\theta}&\left(\frac{\upartial}{\upartial\theta}(u_\theta\sin\theta)+\frac{\upartial u_\phi}{\upartial\phi}\right)=0,
\end{align}
\end{subequations}
where (\ref{dimensional}a,b) are momentum conservation and (\ref{dimensional}c)
is mass conservation; all relative to 
spherical polar coordinates $(r,\theta,\phi)$. In (\ref{dimensional}), $R$ is radius, $h$ is ocean depth ($h\ll R$), $\varOmega$ is rotational frequency, $g$ is surface gravity, $u_\theta$ and $u_\phi$ are respectively colatitude and longitude velocities, $\eta$ is tidal height, $V$ is tidal potential, and $\varPsi$ is perturbation of self-gravity potential caused by $\eta$. The two terms proportional to $2\varOmega$ are Coriolis force due to rotation. We assume that the frictional force of Rayleigh drag is proportional to velocity and the coefficient $\gamma$ is constant and independent of time \citep{ogilvie2009}, i.e. the seafloor frictional coefficient is constant.

Tidal potential has many components \citep{doodson} and the two major components are diurnal and semi-diurnal, of which the latter is stronger. In this paper we focus on the semi-diurnal potential \citep{souchay}
\begin{equation}\label{potential}
V=\frac{1}{4}gR\frac{m_2}{m_1}\left(\frac{R}{d}\right)^{\!3}\left(\cos^2\delta\right)\mathrm P_{2}^{2}(\cos\theta)\cos(2\phi-2\omega t),
\end{equation}
in which $m_1$ and $m_2$ are the mass of respectively Earth and its companion, i.e. Moon or Sun, $d$ is the orbital semi-major axis, $\delta$ is companion's declination, and $\omega=\omega_o-\varOmega$ is tidal frequency in the frame rotating with Earth, where $\omega_o$ is the orbital frequency. $\mathrm P_2^2$ is the associated Legendre polynomial. 

The variables are expanded with spherical harmonics and the spectral coefficients are denoted by hat, e.g. 
\begin{equation}\label{eta}
\eta=\mathrm{Re}\biggl\{\sum_n\hat\eta \mathrm P_n^2(\cos\theta)\,{\mathrm e}^{2{\mathrm i}(\phi-\omega t)}\biggr\},
\end{equation}
where $n$ is the colatitude wavenumber and $\mathrm{Re}$ denotes the real part of a complex variable. The perturbation of self-gravity potential is related to tidal height through
\begin{equation}\label{psi}
\hat\varPsi=\frac{3}{5}\frac{\rho}{\bar\rho}g\hat\eta,
\end{equation}
where $\rho$ is the water density and $\bar\rho$ is the Earth's mean density. The Coriolis force couples different colatitude wavenumbers and numerical calculations are inevitable. As stated before, in the regime of $\varOmega\ll\sqrt{g/R}$ the asymptotic analysis is tractable \citep{longuet-higgins}. However, in the present study, we will investigate a wide regime of Earth's rotation such that we have to perform numerical calculations. The details of numerical method can be found in \citet{wei2019}.

\section{Results}\label{results}

Tidal height $\eta$ is a travelling wave in longitude and its amplitude is a function of colatitude $\theta$. What we will output is the tidal peak $\eta_{\rm max}$, i.e. the maximum of tidal height. The parameters are as follows (from Wiki and NOAA): Earth's mean radius $R=6371$~km, ocean mean depth $h=3688$~m, the density ratio $\bar\rho/\rho=5.514$, surface gravity $g=9.807\,{\rm m/s^2}$, orbital semi-major axis $d=384399$~km for Moon and $1.496\times 10^8$~km for Sun, Earth's mass $m_1=5.97237\times 10^{24}$~kg, companion's mass $m_2=7.342\times 10^{22}$~kg for Moon and $1.9885\times 10^{30}$~kg for Sun, declination $\delta=0^\circ$ corresponding to tidal peak, orbital period $T_o= 720$~hours (30~days) for Moon and $8640$~hours (365~days) for Sun, and period of Earth's rotation $T$ varies on the geologic timescale.

In the first place we need to calibrate the frictional coefficient $\gamma$ with comparison to observation. We calculate the equilibrium tide. By setting $v_\theta=v_\phi=0$ in \eqref{dimensional} and using \eqref{potential}, \eqref{eta} and \eqref{psi} for $n=2$ we obtain
\begin{equation}
\hat\eta=\frac{1}{4}R\frac{m_2}{m_1}\left(\frac{R}{d}\right)^{\!3}\Bigg/\biggl(1-\frac{3}{5}\frac{\rho}{\bar\rho}\biggr),
\end{equation}
and hence
\begin{equation}
\eta_{\rm max}=\frac{3}{4}R\frac{m_2}{m_1}\left(\frac{R}{d}\right)^{\!3}\Bigg/\biggl(1-\frac{3}{5}\frac{\rho}{\bar\rho}\biggr).
\end{equation}
Inserting the numerics of Earth, Moon and Sun, we can obtain $\eta_{\rm max}=0.300$~m for lunar tide and $\eta_{\rm max}=0.138$~m for solar tide, as shown in Table \ref{data}.
Then we calculate the dynamical tide with different $\gamma$ at $T=24$~hours. Table \ref{data} shows the results. At $\gamma=10^{-1}\sqrt{g/R}$ dynamical tide is highly damped. At $\gamma=10^{-2}\sqrt{g/R}$ dynamical tide is close to equilibrium tide. At $\gamma\le 10^{-3}\sqrt{g/R}$ dynamical tide tends to converge and is higher than equilibrium tide. The observation shows that the actual tide is indeed higher than equilibrium tide \citep{wahr}. Thus, in the following calculations we choose $\gamma=10^{-3}\sqrt{g/R}$ to reach the realistic situation, which corresponds to a longer timescale of 224 hours than the early study \citet{webb1980}.

\begin{figure}
\centering
\subfigure[Lunar tide versus Earth's rotation period]{\includegraphics[scale=0.35]{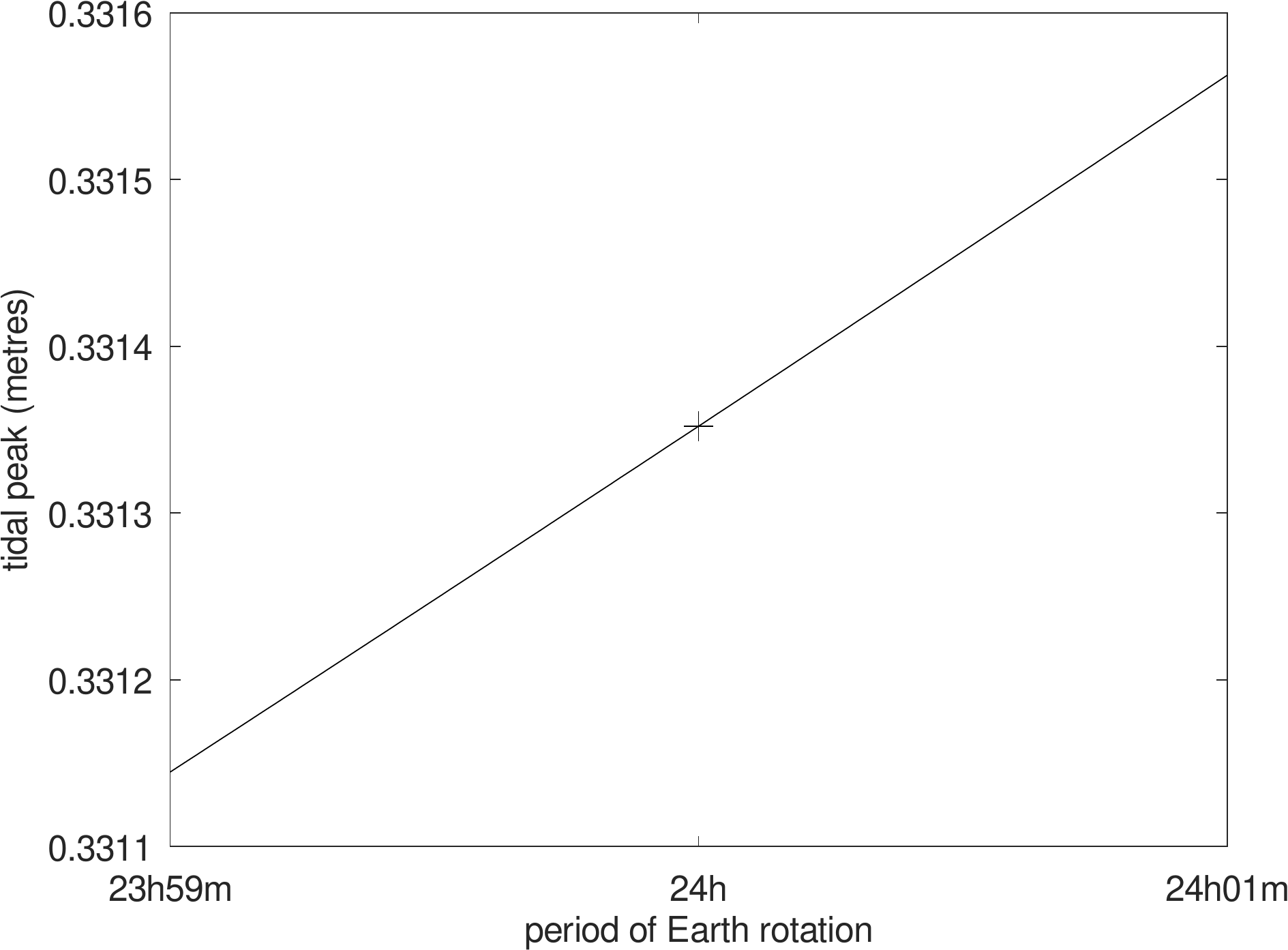}\label{earth1a}}
\subfigure[Solar tide versus Earth's rotation period]{\includegraphics[scale=0.35]{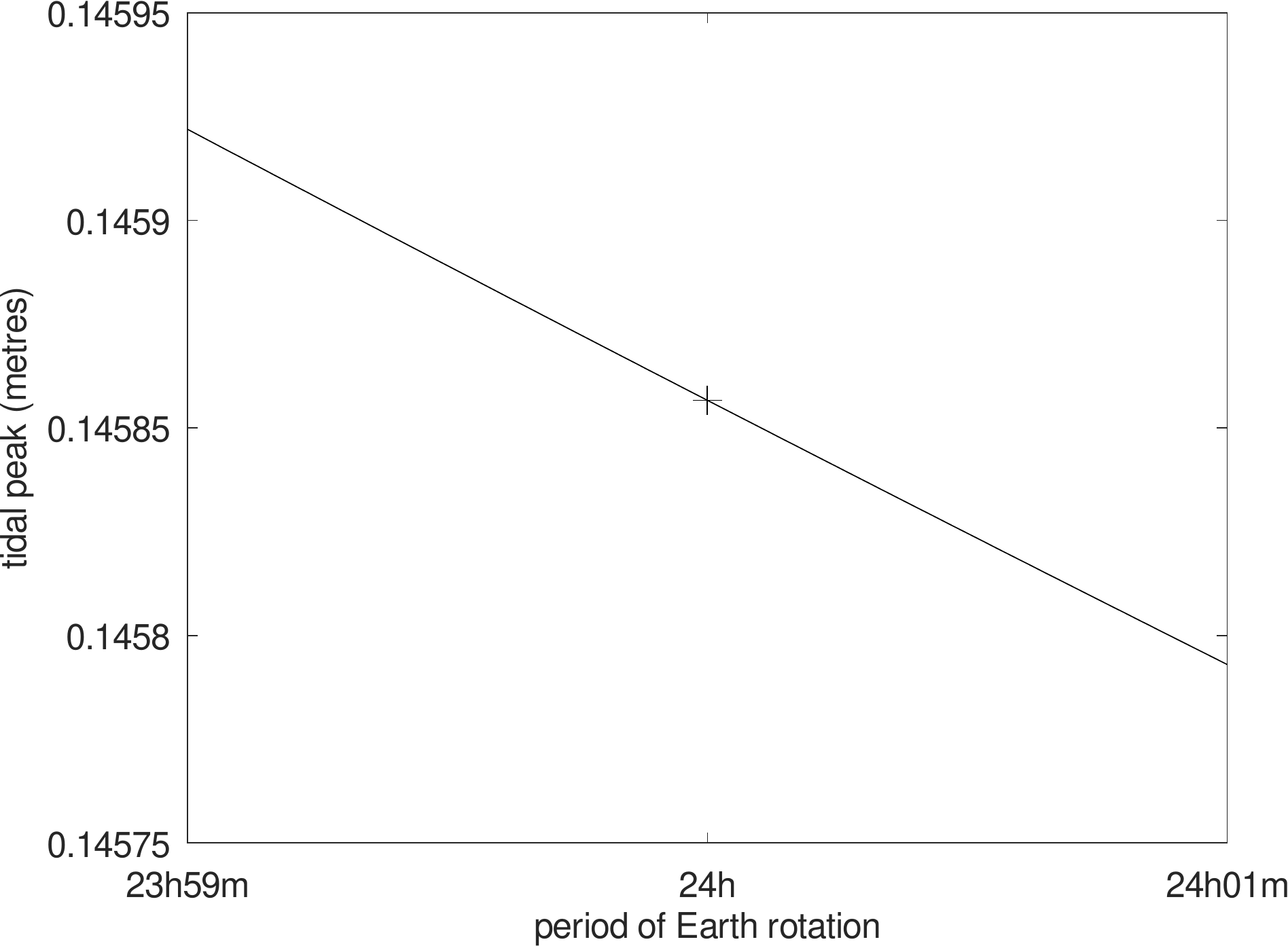}\label{earth1b}}
\caption{Tides on the short timescale of Earth's rotation. Crossings denote the present Earth at $T$ = 24~hours. The ocean depth $h=3688$~m.}\label{fig1}
\end{figure}

Next we study the evolution of tides on the short timescale of Earth's rotation from $23\,$h~$59\,$m to $24\,$h~$01\,$m, i.e. 2-minute variation of Earth's rotation, with the orbital frequency fixed. Figure \ref{fig1} shows the results. Lunar tide increases with the increasing rotational period, which is consistent with the observation of Moon's recession rate \citep{williams}. However, solar tide decreases with the increasing rotational period. Both lunar and solar tides on such a short timescale vary almost linearly with period of Earth's rotation, and $\Delta\eta/\Delta T$ is
\begin{equation*}
\begin{aligned}
0.209 & \mbox{ millimetre per minute for lunar tide}, \\
-0.064 & \mbox{ millimetre per minute for solar tide}.
\end{aligned}
\end{equation*}

\begin{figure}
\centering
\includegraphics[scale=0.7]{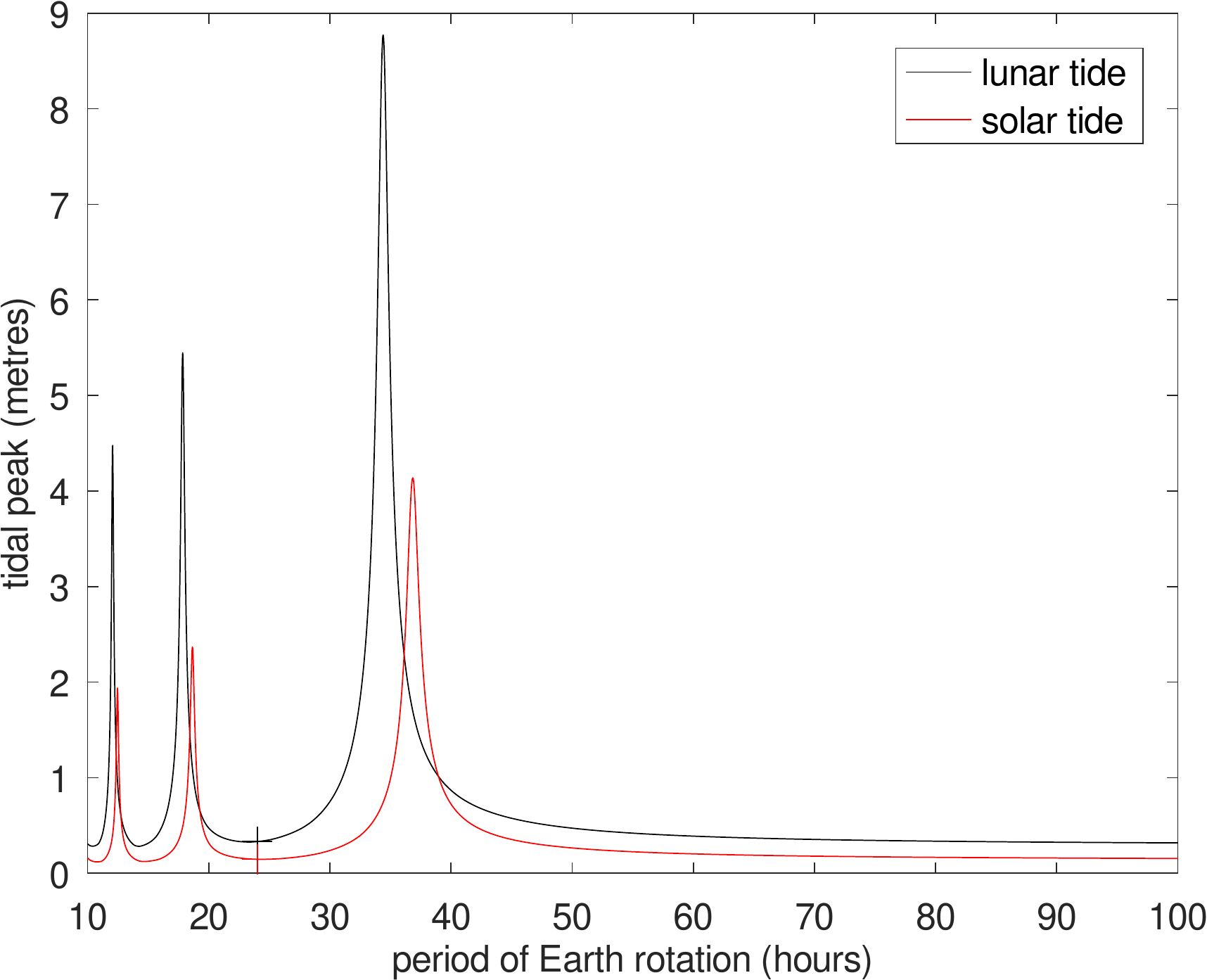}
\caption{Tides on the long timescale of Earth's rotation. Crossings denote the present Earth at $T= 24$~hours. The ocean depth $h=3688$~m. The orbital frequency is fixed. Black curve denotes lunar tide and red curve solar tide. (Colour online)}\label{fig2}
\end{figure}

Now we study the evolution of tides on the long timescale of Earth's rotation. Figure \ref{fig2} shows tides for period of Earth's rotation from 10~hours to 100~hours. On such a long timescale three tidal resonances are found for both lunar and solar tides. Among three resonances next one is higher than last one. The present Earth denoted by crossings is indeed in transition from the previous resonance to the future one. Lunar tide arrives at resonances earlier than solar tide and its resonance peaks are higher than solar tide. Lunar tide experienced the last resonance 5.446~metres at period of Earth's rotation $T=17\,$h~$51\,$m, and will experience the next resonance 8.772 metres at $T=34\,$h~$24\,$m. Solar tide experienced the last resonance 2.368~metres at $T=18\,$h~$39\,$m and will experience the next resonance 4.138 metres at $T=36\,$h~$51\,$m. Tidal resonances will not occur after $T\approx 50\,$h.

This result indicates that our planet once experienced a high global tide and will experience a higher global tide. Local geography can enhance tide, and consequently, during the course of tidal resonances the local tide along continental shorelines can reach very high levels. This could be the cause of great paleofloods and will induce a great flood in the future.

\begin{figure}
\centering
\includegraphics[scale=0.7]{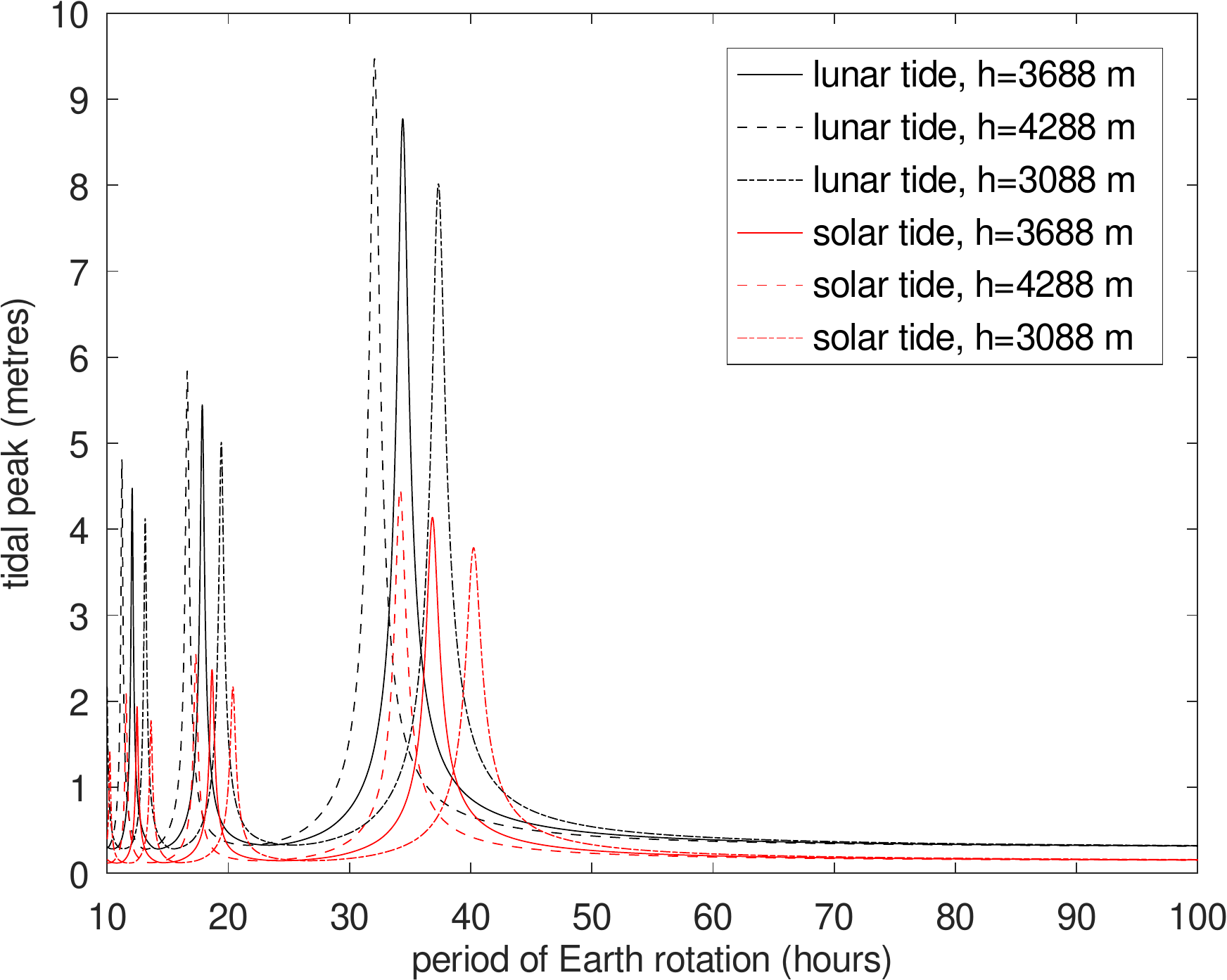}
\caption{Tides on the long timescale of Earth's rotation with the three ocean depths. The orbital frequency is fixed. Black curve denotes lunar tide and red curve solar tide. Solid line denotes the present depth $h=3688$~m, dashed lines $h=4288$~m and dash-dot lines $h=3088$~m. (Colour online)}\label{fig3}
\end{figure}

The thickness of fluid layer, namely ocean depth, influences the natural frequency of surface gravity wave and hence tidal resonances. On a long geologic timescale ocean depth changes. According to \citet{hallam} ocean in the history was about 600 metres higher. In the future, ocean can be shallower because of evaporation. We do not know the accurate change of ocean depth and so we assume the change of $\pm 600$ metres as reference. Figure \ref{fig3} shows the evolution of tides on the long timescale of Earth's rotation with the three ocean depths for comparison with the present depth 3688 metres. It shows that for both lunar and solar tides a deeper ocean advances resonance while a shallower ocean postpones resonance. As stated, the tidal wave is a mixed mode of inertial wave and surface gravity wave. The frequency of the latter is scaled by its phase speed $\sqrt{gh}$, and therefore, a deeper ocean corresponds to a higher frequency of surface gravity wave and hence a higher eigenfrequency of tidal wave, as shown in figure \ref{fig3}. However, the change of ocean depth does not radically change tidal resonances, i.e. it cannot strongly suppress tidal resonances.

\begin{table}
\centering
\begin{tabular}{l|l|lllll|l}
           & equi. & $\gamma=10^{-1}$ & $10^{-2}$ & $10^{-3}$ & $10^{-4}$ & $10^{-5}$ & obs. \\ \hline
lunar      & 0.300 & 0.0594           & 0.305     & 0.331     & 0.332     & 0.332     & 0.385 \\
solar      & 0.138 & 0.0261           & 0.134     & 0.146     & 0.146     & 0.146     & 0.179
\end{tabular}
\caption{Equilibrium tide, dynamical tide with different friction $\gamma$, and observation. Rotational period to calculate equilibrium tide and dynamical tide is 24~hours. Unit of tidal peak is metre. Unit of $\gamma$ is $\sqrt{g/R}\approx 1.241\times 10^{-3}$~s$^{-1}$.}\label{data}
\end{table}

\section{Summary}\label{summary}

In this paper we numerically study the evolution of lunar and solar tides on short and long timescales of Earth's rotation. On the short timescale the change rate of tidal height versus Earth's rotation is obtained. On the long timescale three resonances are found. The previous tidal resonance of 5 metres and the future tidal resonance of 8 metres are both much higher than the present tide of 1~ eter. The two tidal resonances could have caused and will cause great global floodings.

The current theory to calculate the tidal torque and the Earth-Moon orbital evolution is based on equilibrium tide \citep{macdonald,goldreich}. Dynamical tide changes greatly on a long geologic timescale, and is much stronger than equilibrium tide during tidal resonances. The calculation based on equilibrium tide should be revised and tidal resonances should be considered for orbital evolution. Consequently, the tidal torque is stronger than that calculated with equilibrium tide, and both the orbital separation and the slowdown of Earth's rotation should be faster than expected before.

In the present study we focus on tidal resonance but do not consider the interaction of orbital dynamics and tide, i.e. orbit determines tidal force in terms of its strength and frequency and tidal torque determines orbital frequency. This mutual effect will be studied in the presence of tidal resonance. Moreover, we investigate only the linear tide. The nonlinear effect usually suppresses tidal amplitude \citep{wei2016b}. To study linear tide we use the eigenvalue solver but to study the nonlinear effect the time-stepping numerical calculations are necessary, which are more difficult.

\section*{Acknowledgements}
Xing Wei is supported by National Natural Science Foundation of China (grant no. 11872246) and Beijing Natural Science Foundation (grant no. 1202015).

\bibliographystyle{gGAF}
\markboth{\rm X. WEI}{\rm GEOPHYSICAL \&  ASTROPHYSICAL FLUID DYNAMICS}
\bibliography{paper}

\end{document}